\documentclass[modern]{aastex631}
\usepackage[english]{babel}

\usepackage[letterpaper,top=2cm,bottom=2cm,left=3cm,right=3cm,marginparwidth=1.75cm]{geometry}

\usepackage{amsmath}
\usepackage{graphicx}
\usepackage[colorlinks=true]{hyperref}

\newcommand{\beq}{\begin{equation}}
\newcommand{\eeq}{\end{equation}}
\def\sfrac#1#2{{\textstyle{#1\over #2}}}
\newcommand{\bea}{\begin{eqnarray}}
\newcommand{\eea}{\end{eqnarray}}
\newcommand{\nn}{\nonumber}

\begin{document}

\title{No Pulsar Timing Noise from Brownian Motion of the Sun}
\author{James M.\ Cline}
\email{jcline@physics.mcgill.ca}
\affiliation{McGill University Department of Physics \& Trottier Space Institute, 3600 Rue University, Montr\'eal, QC, H3A 2T8, Canada}
\affiliation{CERN, Theoretical Physics Department, Geneva, Switzerland}

\begin{abstract}
It was recently claimed [\cite{Loeb:v4}]
that evidence for a stochastic gravitational wave background
observed by pulsar timing arrays can be attributed instead to random perturbations of the Sun's motion by transiting asteroids.  I show that
that this would lead to a large dipole component accompanying a much smaller quadrupolar perturbation of pulsar timing signals, which would not be confused with a gravitational wave signal. Such an anomalous dipole would have been detected and identified as a spurious background by the PTA collaborations, if it existed.

\end{abstract}

\section*{}

\cite{Loeb:v4} recently argued that the stochastic gravitational wave signal observed by pulsar timing arrays could be
explained by perturbations in the Sun's motion, caused by an unmodeled population of asteroids.  This anomalous motion would
produce a Doppler distortion of the pulsar arrival times, with the same magnitude (at least until \cite{Loeb:v6}; 20\% of the observed signal subsequently) and frequency as the purported gravitational waves.  

Starting in arXiv version 4, \cite{Loeb:v4} recognized that the predicted distortion should be dipolar, whereas the GW signal must be quadrupolar, but minimized this distinction by saying ``it is challenging to separate the 3D random walk of a dipole sourced by a torodial configuration of asteroids around the Sun from a quadrupolar random walk sourced by a stochastic gravitational wave background, given the small number of independent
correlation times ($\sim 5$ periods of $\sim 3$ years) available during the 15 years of PTA
observations.''

We believe this argument is incorrect.  Consider two pulsars with whose directions on the sky relative to Earth are described by the unit vectors
$\hat k_1$ and $\hat k_2$,
such that $\hat k_1\cdot \hat k_2 = \cos\theta$.  
Suppose that there are five sequential asteroids that perturb the Earth's velocity by $\vec v_i$ for $i=1,5$.   The signal from a given pulsar $p$ while the Earth is perturbed by asteroid $a$ is proportional to 
$\cos([\omega_p - \vec v_a\cdot\hat k_p/c]t + \phi_p)$.  The fractional perturbation of the pulsar frequency is given by 
\beq
    {\delta \omega_{p,a}\over \omega_{p,a}} = {\vec v_a\cdot\hat k_p\over c}\,.
\eeq
Since these residuals become stacked during the multiyear observation period, we are interested in $\delta_p = \sum_a \delta 
\omega_{p,a}/\omega_{p,a} \equiv \hat k_p\cdot \vec V$.
Depending on details of the asteroid transit, this sum might be weighted by dimensionless factors $b_a$ so that $\vec V = (1/c)\sum_a  b_a \vec v_a$.
$\vec V$ represents the effective mean velocity perturbation to the Earth during the observing period, in units of $c$.  In \cite{Loeb:2024ekw} it was estimated that $v_a\sim 3\times 10^{-5}$\,cm/s, corresponding to an amplitude $V \sim 10^{-15}$.

Now consider the correlation between timing residuals,
$\langle \delta_1\, \delta_2\rangle$, averaged over all pulsars having similar angular separation $\sim \theta$.  For $N$ pulsars, this is given by 
\bea
    \langle \delta_1\, \delta_2\rangle_\theta &=& 
    {2\over N(N-1)}\sum_{p'<p} {\hat k_p\cdot \vec V}\,
    {\hat k_p\cdot \vec V}\nn\\
        &\times& W(\hat k_p\cdot \hat k_{p'}-\cos\theta)
    \label{avg}
\eea
where $W$ is a window function with support in the angular bin containing $\theta$.  In the limit $N\to\infty$ with pulsars uniformly distributed over the sky, it is straightforward to show that 
$\langle \delta_1\, \delta_2\rangle_\theta\propto \cos\theta$, a purely dipolar distribution.  For finite $N$, we expect fluctuations that can randomly produce a small quadrupole component.

\begin{figure*}[t]
\centerline{
\includegraphics[scale=0.4]{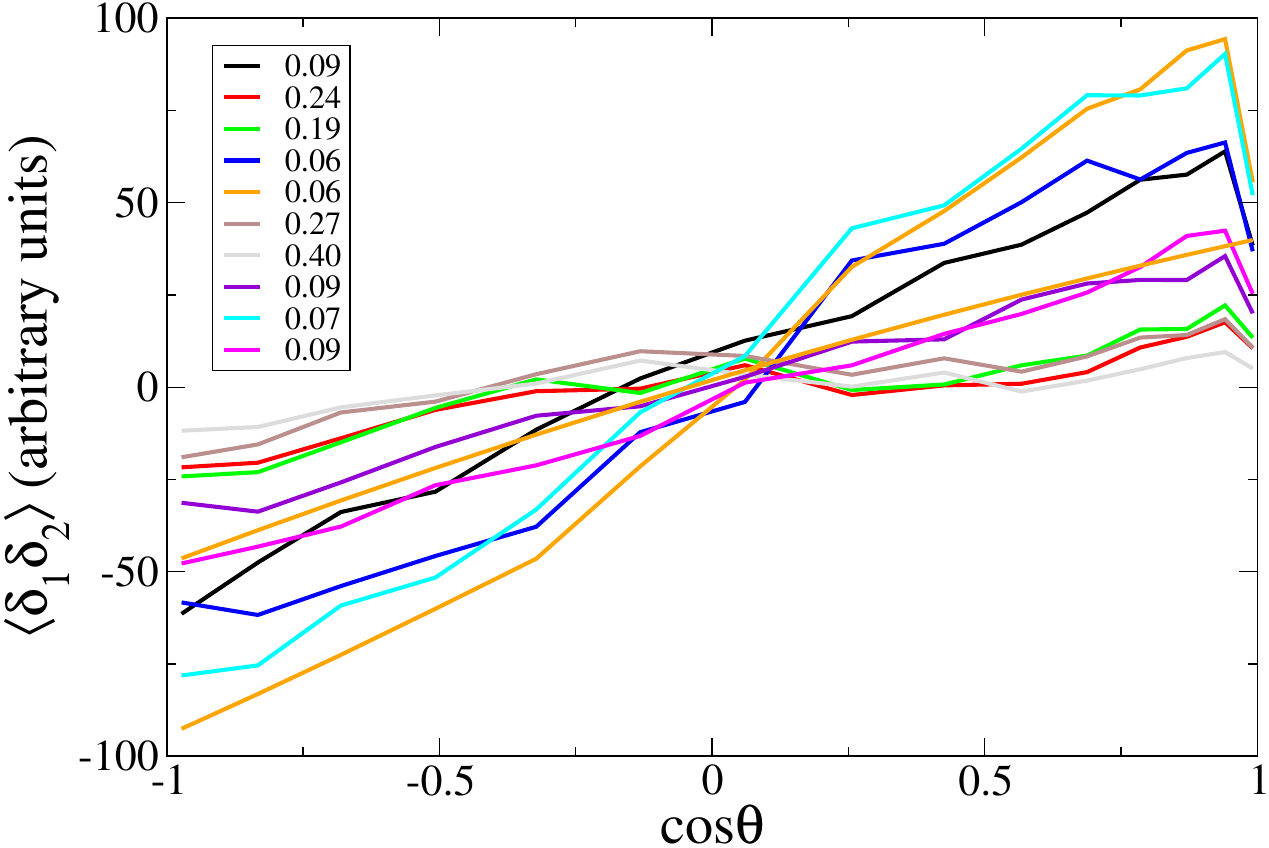}
\includegraphics[scale=0.4]{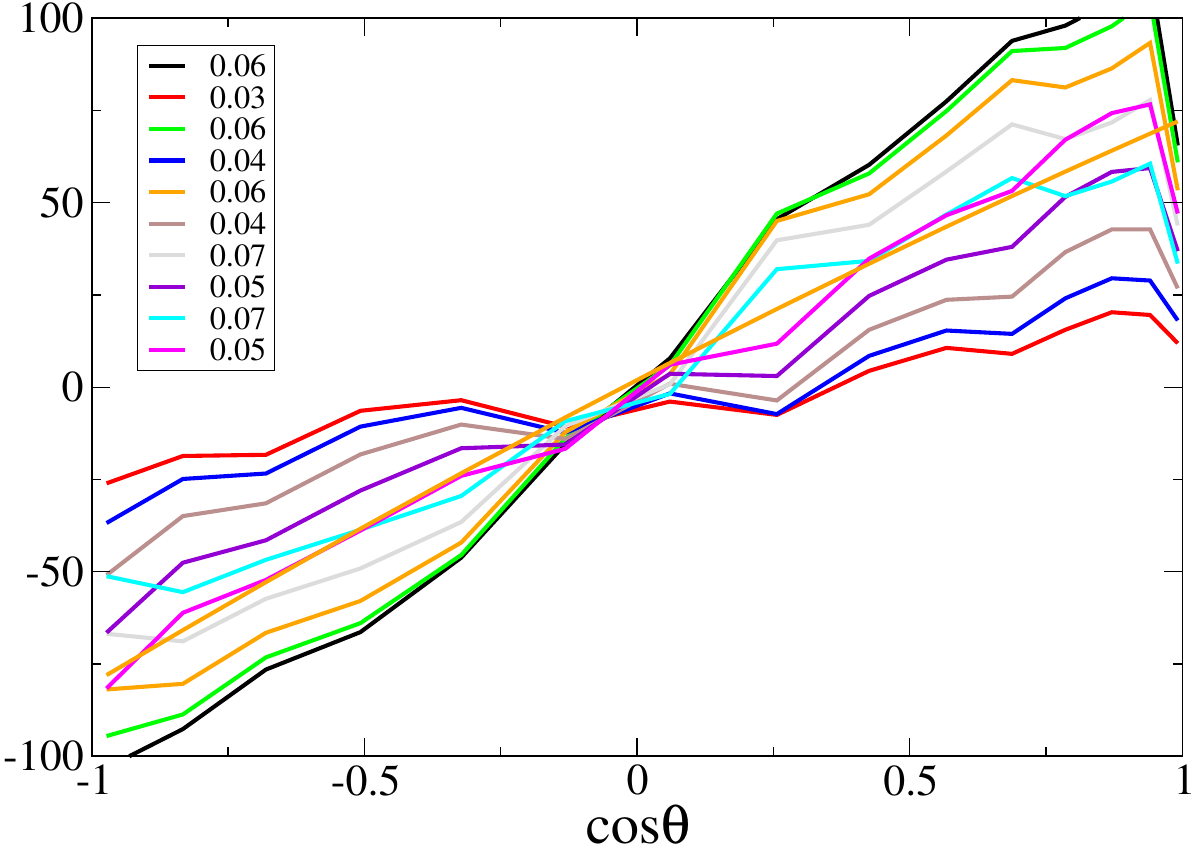}
}
\vspace{-0.2cm}
\caption{Angular correlation function of pulsar timing frequency residuals for random choices of the velocity perturbation $\vec V$.
Left: $\vec V$ is allowed to vary over $4\pi$.  Right: $\vec V$ is restricted to the plane of declination zero, roughly aligned with the ecliptic.  Legend shows relative contribution of quadrupole versus dipole in best fit.}
\label{fig}
\end{figure*}

To quantify this, we have simulated the signal that could be expected for the NANOGrav pulsar timing array [\cite{NANOGrav:2023ygs}], which contains $N=68$ pulsars
with positions on the sky that are identified by their names (right ascension and declination).  Using the same 15 angular bins (each containing $\sim 150$ pulsar pairs) adopted by NANOGrav, the resulting angular correlations are shown for random choices of the velocity
perturbation orientation $\vec V$ in Fig.\ \ref{fig}.  We fit the results to a linear combination of Legendre polynomials $a_1 P_1 + a_2 P_2$, allowing for a dipole plus quadrupole component, and show the ratio $a_2/a_1$.  For velocity perturbations with no restriction on orientation, the average is $\langle a_2/a_1\rangle = 0.16$, while for
$\vec V$ restricted to vanishing declination, as a simplified model for asteroids lying roughly in the ecliptic plane, the average is
$0.05$.

According to \cite{Loeb:2024ekw}, as much as 20\% of the NANOGrav
GW signal could be mimicked by this source of noise; in this case, an accompanying dipolar contamination of intensity $0.2/0.16 \sim 1$ times the quadrupolar signal should have been observed.
The PTA collaborations carefully account for peculiar motion of the Earth in order to remove such backgrounds; hence any unmodeled source strong enough to affect the inferred gravitational wave signal would have stood out in the data as an anomalous dipole contribution [\cite{Caballero:2018lvc,Li_2016,Champion:2010zz}].   The Bayseian odds against the dipole interpretation were found to be $10^7$ in \cite{NANOGrav:2023gor}, contradicting the claim of \cite{Loeb:2024ekw} by many orders of magnitude.  Fig.\ 7 of the NANOGrav discovery paper [\cite{NANOGrav:2023gor}] shows a subdominant contamination of the quadrupole signal by a dipole component at the level of $a_1/a_2 = \sqrt{0.2/1.8} \sim 0.3$, opposite to the behavior in Fig.\ \ref{fig}, which predicts $a_1/a_2 > 6$.
Therefore even a 20\% contribution to the inferred GW signal from anomalous motion of the sun is strongly contradicted by the PTA analysis.

{\bf Acknowledgment.}  I thank Valerie Domcke, Boris Goncharov and Matthew McCullough for helpful discussions.

\bibliography{sample}

\begin{thebibliography}{}
\expandafter\ifx\csname natexlab\endcsname\relax\def\natexlab#1{#1}\fi
\providecommand{\url}[1]{\href{#1}{#1}}
\providecommand{\dodoi}[1]{doi:~\href{http://doi.org/#1}{\nolinkurl{#1}}}
\providecommand{\doeprint}[1]{\href{http://ascl.net/#1}{\nolinkurl{http://ascl.net/#1}}}
\providecommand{\doarXiv}[1]{\href{https://arxiv.org/abs/#1}{\nolinkurl{https://arxiv.org/abs/#1}}}

\bibitem[{Agazie {et~al.}(2023)}]{NANOGrav:2023gor}
Agazie, G., {et~al.} 2023, Astrophys. J. Lett., 951, L8, \dodoi{10.3847/2041-8213/acdac6}

\bibitem[{Agazie {et~al.}(2024)}]{NANOGrav:2023ygs}
---. 2024, Astrophys. J. Lett., 964, L14, \dodoi{10.3847/2041-8213/ad2a51}

\bibitem[{Caballero {et~al.}(2018)}]{Caballero:2018lvc}
Caballero, R.~N., {et~al.} 2018, Mon. Not. Roy. Astron. Soc., 481, 5501, \dodoi{10.1093/mnras/sty2632}

\bibitem[{Champion {et~al.}(2010)}]{Champion:2010zz}
Champion, D.~J., {et~al.} 2010, Astrophys. J. Lett., 720, L201, \dodoi{10.1088/2041-8205/720/2/L201}

\bibitem[{Li {et~al.}(2016)Li, Guo, \& Wang}]{Li_2016}
Li, L., Guo, L., \& Wang, G.-L. 2016, Research in Astronomy and Astrophysics, 16, 006, \dodoi{10.1088/1674-4527/16/4/058}

\bibitem[{Loeb(2024{\natexlab{a}})}]{Loeb:v4}
Loeb, A. 2024{\natexlab{a}}.
\newblock \doarXiv{2405.05410v4}

\bibitem[{Loeb(2024{\natexlab{b}})}]{Loeb:v6}
---. 2024{\natexlab{b}}.
\newblock \doarXiv{2405.05410v6}

\bibitem[{Loeb(2024{\natexlab{c}})}]{Loeb:2024ekw}
---. 2024{\natexlab{c}}, Astrophys. J. Lett., 968, L27, \dodoi{10.3847/2041-8213/ad53c9}

\end{thebibliography}

\end{document}